\documentclass[a4paper]{article}
\usepackage{hiph-art}
\usepackage{graphicx}
\usepackage{amssymb}
\volnumber{19} \issuenumber{1} \edyear{2004}                             
\frompage{000} \topage{000}                                              
\recrevdate{22 April 2004}                                               
\def\rightmark{Lead-glass detector for NA49}
\def\leftmark{F. Sikl\'er} 

\def\pt{$p_T$~}
\def\p0{$\pi^0$}
 
\title{Lead-glass detector for NA49} 
\authors{ 
{Ferenc Sikl\'er $^1$ %
\index{Sikl\'er, F.} %
}\\[2.812mm]
{\normalsize
\hspace*{-8pt}$^1$ RMKI,\\ 
1211 Budapest, Hungary\\[0.2ex]
}}
 
\abstract{An experimental report on the construction and operation of a
lead-glass calorimeter at the CERN-NA49 experiment is presented.}


 
\makeindex
\begin{document}
 
\maketitle

\section{Introduction}

After the exciting and interesting RHIC results on suppression of high \pt
particles in very high energy nucleus-nucleus collisions -- and their
absence for proton-nucleus reactions -- it is reasonable to ask: what
about lower, SPS, energies, do we see anything interesting there?

In case of proton-nucleus collisions the enhancement of high \pt particles
compared to nucleon-nucleon collisions is the well known Cronin-effect.
Some recent calculation of this phenomenon for SPS energy is given in
Ref.~\cite{Barnafoldi:2003kb}. As the data available from experiments are
scarce (e.g. unpublished analysis from WA98), it is worthwhile to measure
high \pt particles, for example neutral pions, with a small supplement
using the existing detector system of the NA49 experiment. Its completion
can also help when making exclusive studies, because the experiment up to
now could detect charged particles only, with the exception of neutrons.

\begin{figure}
 \includegraphics[width=\linewidth]{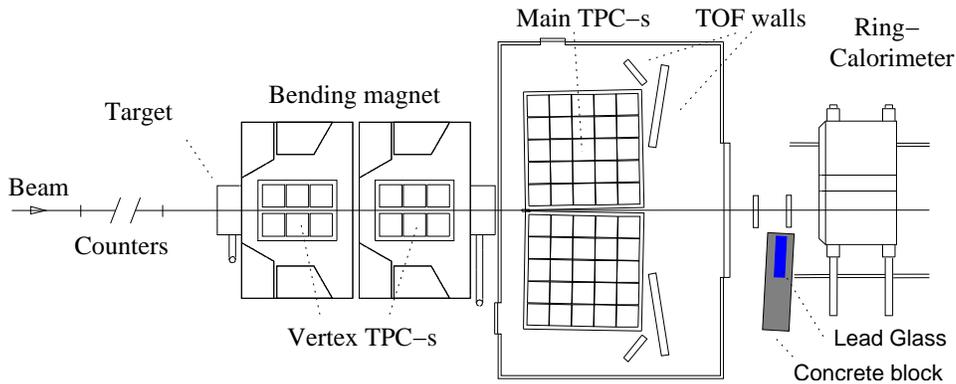}

\caption[]{Layout of the NA49 experiment with the position of the new
lead-glass detector.}

\label{fig:layout}
\end{figure}


The NA49 experiment is a large acceptance hadron detector for charged
particles (Fig.~\ref{fig:layout}). Although the SPS is closed, our group
got 10 days of running in 2003, in order to test a lead-glass detector
prototype, put behind the time projection chambers. For most of the
particles having photonic decay one has acceptance in the central region.


%
%
%

\section{OPAL end cap electromagnetic calorimeter}

The hardware for the new detector came from the OPAL experiment which
finished data-taking in 2000. During the dismantling big part of the end
cap calorimeter hardware and electronics \cite{Jeffreys:1989yc} was
salvaged. The calorimeter functions in the following way. The incoming
$\gamma$ creates an electromagnetic shower, which produces Cherenkov
light. The lead-glass has good light transmission below 400 nm and it is
long enough, thus a shower is easily contained. The light is converted to
electronic signal, by the vacuum photo triode, and it is further amplified.
The triode contains one dynode only, meaning small amplification. The unit
is sensitive to magnetic fields but also tolerates high voltage changes.
The device has acceptable resolution in the percent range and enables
hadron-electron separation.

The resulting negative signals are integrated by a dual 12 bit charge
integrating ADC of the type CIAFB F583C \cite{Beer:ym}. It digitizes 96
channels, measuring both the signal and the amplified one, thus achieving
15 bit dynamic range. The device can be gated from the front, but the
generation of a test gate and a test pulse is also possible. The device
gives out the analog trigger output before conversion, thus it may be used
for triggering purposes. To enable fast clear, the conversion can be
delayed by 10-60 $\mu$s, the conversion time is 1 ms for 96 channels. The
device was used with 2 $\mu$s gate and read out via its Fastbus interface.

\begin{figure}
 \includegraphics[width=0.45\linewidth]{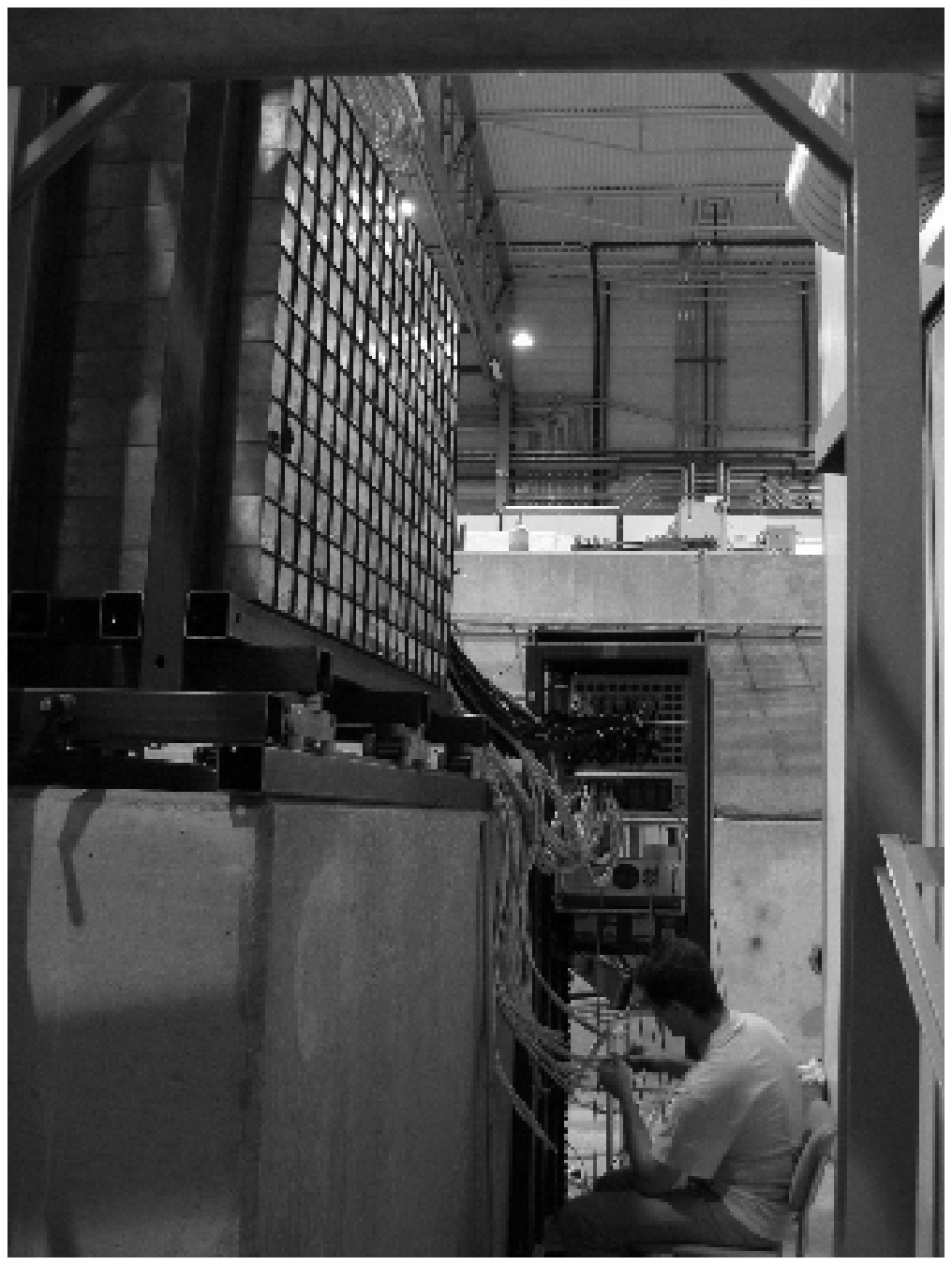}
 \hspace{0.1in}
 \includegraphics[width=0.45\linewidth]{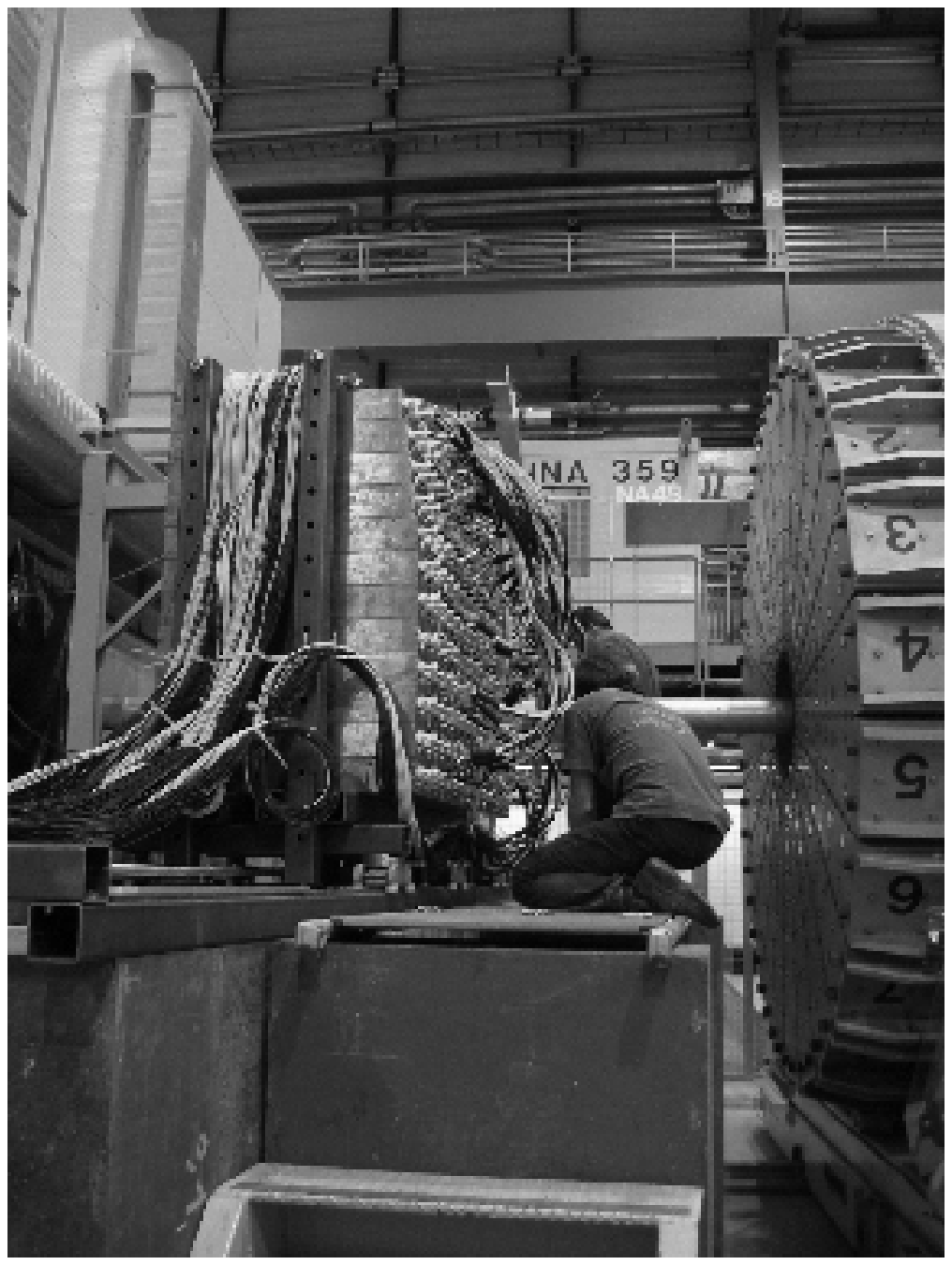}
 
 \caption{\label{fig:install} Installation and assembly. Front of the
detector is shown on the left, with the in-area electronics. The back side
and cabling can be seen one the right.}

\end{figure}

\section{Assembly}

A lead-glass wall with 16$\times$12 units was assembled, giving 1.5 $m^2$
sensitive area. Big part of electronics was placed in the experimental
area, see photos in Fig.~\ref{fig:install}.

In order to enhance events with photons, a photon-trigger has been
developed and built by the institute: electronic cards sum the signals
from four adjacent channels. The trigger thresholds can be set one-by-one
via serial port, using a graphical interface. The response time of the
trigger is of the order of 0.5 $\mu$s, depending on signal amplitude.

The data acquisition software runs on FIC 8234 machine, with Motorola
68040, running OS-9 (interrupt handling, events, modules, semaphores).
Fastbus is reached via FVSBI interface. The software not only controls the
measurement, but provides an on-line display of the occurring hits in the
detector.

\begin{figure}
 \includegraphics[width=0.45\linewidth]{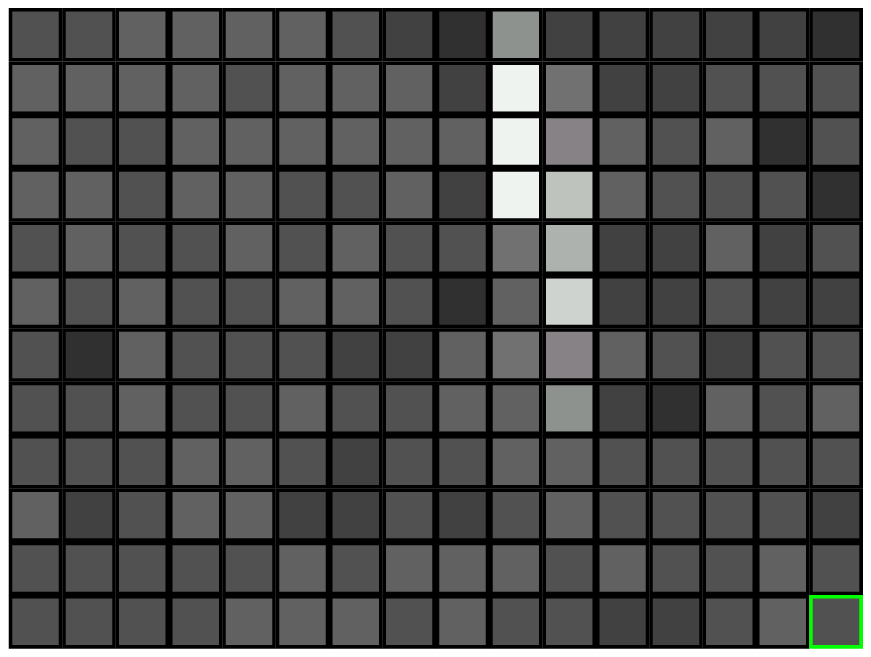}
 \hspace{0.1in}
 \includegraphics[width=0.45\linewidth]{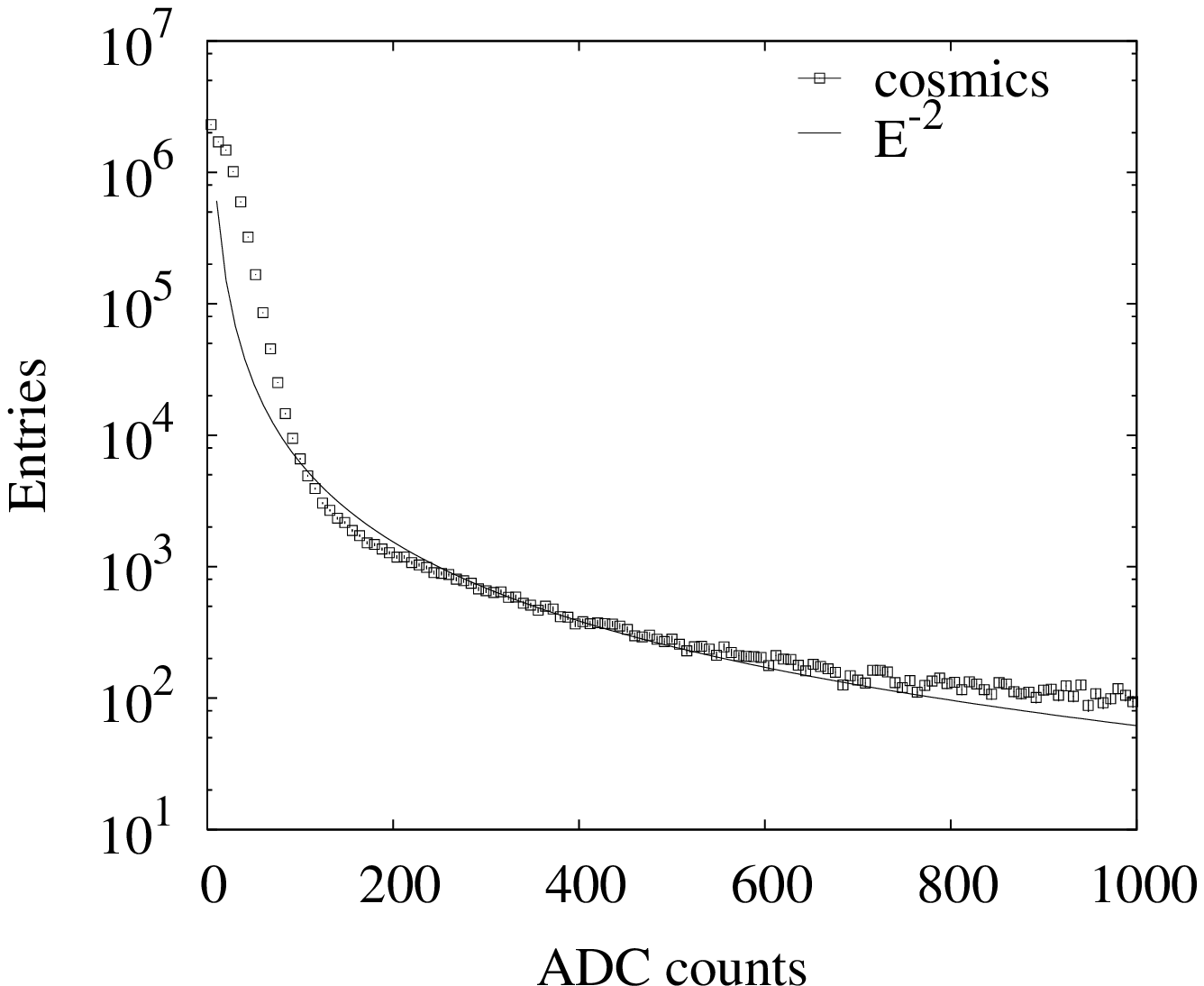}
 \vspace{0.1in}

 \caption{Left: example of a cosmic event, creating vertical shower
starting at the top of the detector. Cells with higher energy deposit are
whiter. Right: distribution of ADC counts in all cells, compared to the
expected $1/E^2$ shape.}

\label{fig:result}
\end{figure}

\section{Data taking}

As a first check, data have been taken with triggering on cosmic particles
(Fig.~\ref{fig:result}). The obtained uncalibrated ADC spectrum, which in
this sense is the energy spectrum, gives dependence close to $1/E^2$. By
plotting the units separately one finds that the gains are within
20-30\%.

Due to problems with the accelerator and some chambers, half of the
beam-time was lost. 
Finally interactions of fragmented deuterons on liquid hydrogen
target with spectator proton trigger, n+p reactions, were taken. The
photon-trigger, mostly giving high \pt {\p0}s, appeared to be too slow:
only minimum bias events were recorded. Still, this data sample was enough
for checking hit frequencies, event multiplicities.

The first attempt on reconstructing {\p0}s failed, the correct
determination of pedestal and relative gain appears to be crucial. This
information can be extracted from the acquired data and the study of
events with cosmics. Nevertheless an absolute calibration with electron
beam would be important.

\section{Summary}

Theoretical predictions and poor measurements of high \pt particles at SPS
show that the project discussed above is reasonable. Using existing
hardware parts a working detector could be built which provided results
only from some days of running.

\vspace{-0.05in}
\section*{Acknowledgment}

Full details of the analysis, with calibrated units and nice \p0 mass
spectrum, will be available in the thesis of A.~L\'aszl\'o. This would not
have been possible without the work of G.~Vesztergombi and D.~Varga, many
graduate and PhD students, with other researchers of the institute. This
work is supported by the Hungarian Sientific Research Fund (T043514,
F034707). The author wishes to thank to the J\'anos Bolyai Research Grant.
 

\vfill\eject

\begin{thebibliography}{99}  

\bibitem{Barnafoldi:2003kb}
G.~G.~Barnafoldi, G.~Papp, P.~Levai and G.~I.~Fai,
arXiv:nucl-th/0307062.

\bibitem{Jeffreys:1989yc}
P.~W.~Jeffreys {\it et al.},
Using
Nucl.\ Instrum.\ Meth.\ A {\bf 290} (1990) 76.

\bibitem{Beer:ym}
A.~Beer, G.~Critin and G.~Schuler,
Density
Nucl.\ Instrum.\ Meth.\ A {\bf 234} (1985) 294.
 
\end{thebibliography}
\end{document}